# STATE-OF-THE-ART IN EMPIRICAL VALIDATION OF SOFTWARE METRICS FOR FAULT PRONENESS PREDICTION: SYSTEMATIC REVIEW

Bassey Isong[1] and Obeten Ekabua[2]

[1]Department of Computer Sciences, North-West University, Mmabatho, South Africa
[2]Department of Computer Science, Delta State University, Abraka, Nigeria

## ABSTRACT

*With the sharp rise in software dependability and failure cost, high quality has been in great demand. However, guaranteeing high quality in software systems which have grown in size and complexity coupled with the constraints imposed on their development has become increasingly difficult, time and resource consuming activity. Consequently, it becomes inevitable to deliver software that have no serious faults. In this case, object-oriented (OO) products being the de facto standard of software development with their unique features could have some faults that are hard to find or pinpoint the impacts of changes. The earlier faults are identified, found and fixed, the lesser the costs and the higher the quality. To assess product quality, software metrics are used. Many OO metrics have been proposed and developed. Furthermore, many empirical studies have validated metrics and class fault proneness (FP) relationship. The challenge is which metrics are related to class FP and what activities are performed. Therefore, this study bring together the state-of-the-art in fault prediction of FP that utilizes CK and size metrics. We conducted a systematic literature review over relevant published empirical validation articles. The results obtained are analysed and presented. It indicates that 29 relevant empirical studies exist and measures such as complexity, coupling and size were found to be strongly related to FP.*

## 1. INTRODUCTION

In today's e-world, the importance of software technologies have been seen in different kinds of productsand services used in everyday life. The exponential growth of software dependability poses the demand for high quality from users and to meet this demand, today software has grown in size and complexity [1][2][3][4]. This is because quality of software is the key determinant of the success or failure of an organization [5]. However, guaranteeing high quality in this modern age of large software systems development, increased difficulty, time and resource consumption has become the order of the activity [4][5][6]. Given the size, the complexity and the constraints imposed on the development, it is inevitable to deliver to customers software that have no faults [1][3][4]. In particular, object-oriented (OO) products with its unique features could have introduced some faults that are hard if not impossible to find or pinpoint change impacts during maintenance. Faults in software are errors introduced during the software development activity that can lead software to fail or not meeting customers' expectations. Though, it is difficult to find and fix faults before product release, the earlier this is done the, lower the costs and the higher the product quality would be [1][5][7][8][9][10][11] [12][13][14][15]. In software engineering, one way to assure software quality cost-effectively is the use of software metrics.

Software metrics usage during development process, especially at the early phases is critical to ensuring high quality in the software systems. There are used as a tool in software organizations to assess software quality, monitor, control and take useful managerial and technical decisions aimed at improving the software [16][17][18]. Existing software metrics are broadly classified into traditional metrics and OO metrics [18]. Moreover, many OO metrics have been proposed and developed for assessing OO design and codes quality [1][2][3][6][7][8][17][19][22]. OO product metrics capture different software attributes such as class complexity, inheritance, couplings and cohesions [10][16]. These structural properties are used to determine products quality and complexity [10][19]. One of such OO metrics is the CK metric suit [16].

Albeit a greater amount of software faults found in software applications today are believed to concentrate only on few classes of the system [10][20], what is more important is when such faults are identified. In the world of OO systems, one viable approach used by engineers is to identify faulty OO classes during the software development early stage through the construction of quality models for prediction utilizing OO metrics and historical measures [1][3][4][6][10][11][22][31]. The construction of these models can be used by organizations in the identification of possible classes which are faulty either in the future applications or release and to identify where resources are needed most [10]. Thus, it assist organizations to focus quality improvement activities, make decisions, plan and schedule development activities in order to produce high quality product within time and budget [10][19][21]. For instance, testing large systems today is complex and time-consuming activity [5][6][10]. Therefore, predicting faulty components early would allow organizations to take actions aim at mitigating against the high risk posed by the faults which are likely to cause failure in the field. Such activities include focusing testing and verification resources on such classes to avoid rework that could be costly [10].

However, for OO design metrics to accurately predict faults in OO classes there have to be empirically validated. That is, establishing which metrics are related to important external quality attributes like class fault-proneness (FP). The essence is that, OO metrics are of no or little value if such relationship is not empirically validated [10]. Nevertheless, few empirical validation studies exist that have validated or re-validated OO metrics with respect to FP [2][3][6][7][8][17][19][22][23][24]. In addition, these studies proposed and developed several prediction models that make use of FP and OO metrics as dependent and independent variables respectively. Among such validated OO metrics is the CK metric suite and size metric. Several empirical studies in the literature has shown that some metrics are significantly or insignificantly related to FP [2][3][6][7][8][17][19][22]. Furthermore, their findings appeared not to be consistent [2]. For example, in one study a metric is considered related FP but insignificant related to FP in another study. However, this could affect decision making in choosing directly metrics that are associated with FP of a class. Hence, *which of these metrics are actually related to the FP of a class?*

To establish OO design metrics that are related FP and are generic, this paper performed a systematic literature review (SLR) using published empirical validation studies of CK +SLOC metrics. The basis for this SLR is that the authors lack resources to perform empirical study on real-world software systems, only few SLR on the CK + SLOC point of view exist within periods considered and lastly, to bring together the state-of-the-art in fault prediction using FP and CK + SLOC metrics. The study is specifically designed to assist software engineers take quick decision regarding generic metrics that are suitable for fault prediction in a class when CK+SLOC metrics are used.

The remaining part of the paper is organized as follows: Section 2 is the description of the metrics used in this study, Section 3 is the research method used, Section 4 is analysis, Section 5 is the study discussions and Section 6 is the conclusions.

## 2. METRICS STUDIED

The metrics considered in this study is the CK metric suit and the studies that have empirically validated them. Moreover, the study also consider product size metric known as SLOC due to its strong relationship with FP [1][17][22][25]. These metrics are shown on Table 1 alongside their descriptions. They consist of six (6)OO design metrics and one size metric from the traditional product metric.

Table 1. Metrics studied [16]

| Metric | Definition |
|---|---|
| **CK:** | |
| Weighted Methods per Class (WMC) | A count of methods implemented within a given class. |
| Coupling between Objects (CBO) | CBO for a class is count of the number of other classes to which it is coupled and vice versa. |
| Response for a Class (RFC) | The count of methods implemented within a class plus the number of methods accessible to an object class due to inheritance. |
| Lack of Cohesion (LCOM) | For each data field in a class, the percentage of the methods in the class using that data field; the %s are averaged then subtracted from 100 %. |
| Depth of Inheritance (DIT) | The length of the longest path from a given class to the root in the inheritance hierarchy |
| Number of Children (NOC) | The NOC is the number of immediate subclasses of a class in a hierarchy. |
| **Size:** | |
| Source Lines Of Code (SLOC) | It counts the lines of code (nonblank and non-commented) in the body of a given class and all its methods |

## 3. RESEARCH METHODOLOGY

This study has been conducted by strictly following the guidelines for performing SLR offered by Kitchenham et al [26][39]. SLR is a secondary study which provides the means to gather and analyse a collection of published research findings that assist in answering stated research questions. This SLR results will be useful in identifying the current state-of-the-art of the empirical validation of the relationship between CK metrics, size measure and class FP. The steps involve are discussed as follows.

### 3.1. Research Questions

This study is aim at providing empirical evidences from published studies in the literature to identify which of the CK and SLOC metrics are strongly associated with class FP in terms of significance level. Thus, the research questions intended to be answered are as follows:

*RQ1: Which metric (s) within the CK metric suite and SLOC is related to the FP of a class?*
This question is designed to provide answers on which metrics are significant or not significant with FP of OO classes. This study will limit its findings to significance and insignificance relationship regardless of if the relationship is positive, negative, weak, strong or severe.

*RQ2: What techniques are being used to empirically validate the metrics in RQ1 and which is the best?*
This question will be used to explore the state-of-the-art in fault prediction using FP and CK and SLOC with respect to the statistical techniques, models and variables used.

*RQ3: To what extent have the metrics in RQ1 been validated?*

This question is designed to elicit information about the state-of-the-art in fault prediction using FP and CK and SLOC metrics with respect to the programming language used, the settings of the validation, the type of systems used and the product release used.

*RQ4: Of what relevance are the empirical validations of software metrics?*
This question is designed to provide the relevance of empirically validating the relationship between FP and CK and SLOC metrics

*RQ5: Are there generic software metrics for predicting faulty classes?*
This question is designed to provide answers on whether there exist validated OO metrics which are generic in the prediction of FP of OO software systems. This is important to help developers or managers make quick decisions during software development.

### 3.2. Search Strategy, Terms, Resources and Selection

Search strategy has the goal of ensuring that only relevant studies or articles appears in the search results. In this study, we considered the review of 17-years' efforts in empirical validation of CK and SLOC metrics, between the period of January 1995 to December 2012. These periods were strategically chosen with respect to the birth of CK metric suite and to sufficiently explore the information provided within these periods. To this end, all studies published after the December, 2012 are not included. Another review will be carry out to cover the years after December 2012 in order to enable us perform comparisons on the state-of-the-art in fault predictions with those periods.

However, search results are well documented to enhance the clarity of the search process and avoid duplications. Search terms or strings were formulated and applied manually during the process by following the steps suggested in [26]. For more details, refer to Isong and Ekabua [43]. Furthermore, we limited the search for relevant studies to electronic databases such as Google Scholar, Compendex, Inspec and Scopus. There are subsets of databases largely recognized by researchers worldwide and known to contain relevant journals and conferences articles within computer science and software engineering. Databases such as IEEE Xplorer, Springer Link and ACM were not searched directly since they were indexed or linked to the Engineering Village database (Compendex and Inspec). Based on the study selection criteria designed, relevant studies were selected during the review process to be used for data extraction. This is accomplished by defining basic and detailed inclusion and exclusion criteria in accordance with the research questions. In addition, quality assessment criteria was used to assess the quality of all included studies. This is important to understanding the state of empirical validation of each included study. In this case, each selected study is assessed against a number of checklist questions and each question answered with Yes or No.

### 3.3. Data Extraction and Execution

This study designed data extraction form or template and used for information extraction. All inconsistencies and difficulties encountered were resolved. Moreover, the extracted data was checked at least twice by the authors. To achieve this, the authors independently carry out the process involves in searching for articles that satisfied the defined inclusion and exclusion criteria in the data extraction phase. The databases were scanned using the search terms/strings and the basic defined inclusion and exclusion criterion on the articles to select relevant articles.

With the data extraction forms, each author performed validation on the extracted data in order to accomplish inter-study consistency. All the information about the total number of results obtained (selected and rejected articles) from each database were recorded in the search record. A total of 4683 articles that cited CK and SLOC metrics were retrieved after applying all search terms. At first, studies were excluded after reading their title and abstracts. Furthermore, the remaining studies were selected by applying thorough exclusion and inclusion criteria. More so, the extracted data was compiled and organized quantitatively to answer the stated research questions. Table 2 shows the list of selected Journals and Conferences papers considered in this SLR. Only author's first name was included due to space constraint. For more information, out of the 29 studies selected, 5 are from conference proceedings and 24 are from journals.

Table 2. Selected Articles

| Id | Ref. | Year | Author | Title |
|---|---|---|---|---|
| 1 | [10] | 2001 | Emam et al | The prediction of faulty classes using object-oriented design metrics |
| 2 | [19] | 1998 | Briand et al | A Comprehensive Empirical Validation of Design Measures for OO Systems |
| 3 | [21] | 2001 | Emam et al | The Confounding Effect of Class Size on the Validity of OO Metrics |
| 4 | [17] | 2002 | Yu et al | Predicting FP using OO Metrics: An Industrial Case Study |
| 5 | [1] | 2008 | Zu et al | An Empirical Validation of Object-Oriented Design Metrics for Fault Prediction |
| 6 | [22] | 2000 | Briand et al | Exploring the relationships between design measures and software quality in OO systems |
| 7 | [6] | 1996 | Basili et al | A Validation of Object-Oriented Design Metrics as Quality Indicators |
| 8 | [25] | 2005 | Gyimothy et al | Empirical Validation of OO Metrics on Open Source Software for Fault Prediction |
| 9 | [23] | 2007 | Olague et al | Empirical Validation of Three Software Metrics Suites to Predict FP of OO Classes Developed Using Highly Iterative or Agile Software Development Processes |
| 10 | [3] | 2006 | Zhou et al | Empirical Analysis of Object-Oriented Design Metrics for Predicting High and Low Severity Faults |
| 11 | [4] | 2010 | Singh et al | Empirical validation of object-oriented metrics for predicting FP models |
| 12 | [8] | 1999 | Tang et al | An Empirical Study on Object-Oriented Metrics |
| 13 | [5] | 2003 | Succi et al | Practical assessment of the models for identification of defect-prone classes in OO commercial systems using design metrics |
| 14 | [2] | 2003 | Subramanyam et al | Empirical Analysis of CK Metrics for OO Design Complexity: Implications for Software Defects |
| 15 | [24] | 2009 | Aggarwal et al | Empirical Analysis for Investigating the Effect of Object-Oriented Metrics on FP: A Replicated Case Study |
| 16 | [27] | 2001 | Briand et al | Replicated Case Studies for Investigating Quality Factors in OO Designs |
| 17 | [28] | 2008 | Olague et al | An empirical validation of OO class complexity metrics and their ability to predict error-prone classes in highly iterative, or agile, software: a case study |
| 18 | [11] | 2010 | Malhotra et al | Empirical validation of OO metrics for predicting FP at different severity levels using support vector machines |
| 19 | [29] | 2012 | S. Singh et al | Validating the Effectiveness of OO Metrics over Multiple Releases for Predicting FP |
| 20 | [12] | 2008 | Shatnawi et al | The effectiveness of software metrics in identifying error-prone classes in post-release software evolution process |
| 21 | [9] | 2005 | Janes et al | Identification of defect-prone classes in telecommunication software systems using design metrics |
| 22 | [30] | 2009 | English et al | Fault Detection and Prediction in an Open-Source Software Project |
| 23 | [31] | 2008 | Goel et al | Empirical Investigation of Metrics for Fault Prediction on OO Software |
| 24 | [32] | 2011 | Shaik et al | Investigate the Result of Object Oriented Design Software Metrics on FP in Object Oriented Systems: A Case Study |
| 25 | [33] | 2011 | Dallal, J.A | Transitive-based object-oriented lack-of-cohesion metric |
| 26 | [34] | 2010 | Dallal et al | An object-oriented high-level design-based class cohesion metric |
| 27 | [35] | 2010 | Zhou et al | On the ability of complexity metrics to predict fault-prone classes in Object-Oriented systems |
| 28 | [36] | 2007 | Pai et al | Empirical Analysis of Software Fault Content and FP Using Bayesian Methods |
| 29 | [37] | 2012 | Johari et al | Validation of OO Metrics Using Open Source Software System: An Empirical Study |

# 4. ANALYSIS AND RESULTS

This section presents the analysis of the findings in the SLR by answering the above stated research questions.

## 4.1. CK and SLOC Metrics Relationship with Fault Proneness

*RQ1: Which metric (s) within the CK metric suite and SLOC is related to the FP of a class?* In this study, 29 studies are considered on the basis of empirical validation of software metrics. In these studies, 7 metrics (i.e. 6 CK metrics and 1 "traditional" metric) were empirically validated as related the FP of OO class. However, the analysis carried out shows that some metrics are significant, some strongly significant, some insignificant, while some are negatively significant across the studies. Additionally, some studies categorized their findings in terms of significance and insignificance based on the severity of the fault found such as high, medium, low and ungraded [18][38]. But in this study there is no distinction as to whether a significance is positive or negative and fault severity [3]. Nonetheless, analysis presented in Table 3, 4, 5, 6 and 7 indicates that some metrics are considered to be significant in some studies, insignificant in others while few studies did not measured the metrics. The analysis of the finding is as follows:

*Complexity measure:* For WMC, the validation based on the hypothesis constructed confirms that classes having more member functions or methods are more likely to have faults than classes with small or no member functions. However, 22 studies confirmed WMC significance relationship with the FP of OO classes, one study [2] found considered it to be insignificant while 6 others studies did not consider it their studies.This is captured in Table 3.

Table 3. WMC Validation

| Metric | Significant | Insignificant | N/A |
|---|---|---|---|
| WMC | [10],[40],[17],[1],[22],[6],[25],[23],[3],[4],[8],[2],[24],[27],[11],[29],[12],[31],[32],[35],[36],[37] | [2] | [5],[9],[34],[19],[30],[33] |

**N/A= not applicable

*Coupling measures:* Analysis indicates that 23 of the studies found CBO to be having strong influence on class FP. The significance stems from the fact that a class that is highly coupled tends to be more fault-prone than class that is loosely coupled. To this end, one study found CBO to be insignificant but CBO was not measured in 5 studies.(See Table 4) Moreover, RFC was found to be strongly significant related to class FP in 24 studies. The findings confirms that a class with higher response sets tends to be more fault-prone than others with less response sets. Interestingly, none of the studies found RFC insignificant but 5 of the studies did not measure RFC.

Table 4. CBO and RFC Validation

| Metric | Significant | Insignificant | N/A |
|---|---|---|---|
| CB0 | [10],[19],[21],[17],[1],[22],[6],[25],[23],[3],[4],[2],[24],[27],[11],[29],[12],[9],[30],[31],[32],[36],[37] | [8] | [5],[28],[35],[34],[33] |
| RFC | [10],[19],[21],[17],[1],[22],[6],[11],[23],[3],[4],[8],[5],[24],[27],[11],[29],[12],[9],[30],[31],[32],[36],[37] | - | [2],[28],[35],[34],[33] |

**N/A= not applicable

*Cohesion measure:* Based on the analysis carried out in this study, it shows that 14 studies found LCOM to besignificantly related to class FP. Nevertheless, only 4 studies considered LCOM to be insignificant while 11 studies did not measure LCOM in their study. This is shown in Table 5. The overall results confirmed that a class with low cohesion value is more likely to have faults than class with high cohesion value.

Table 5. LCOM Validation

| Metric | Significant | Insignificant | N/A |
|--------|-------------|---------------|-----|
| LCOM | [19],[21],[17],[25],[23],[3],[4],[24], [27],[11],[9],[32],[35],[34],[37] | [21],[6],[29],[31] | [10],[1],[22],[8],[5],[2], [28],[12],[30],[33],[36] |

**N/A= not applicable

*Inheritance measures:* In the perspective of inheritance measure of a class, results has it that only 9 studies found DIT to be significantly (strong and weak) related to FP. However, about 15 studies considered it to be insignificant while 5 studies did not measure it. With emphasis on the insignificance of DIT, it indicates that a class with higher number of inheritance hierarchy is not likely to have faults. Furthermore, only 3 studies found NOC to be significantly related to FP while 15 studies considered it insignificant. With the insignificance results, it show that a class having a higher number of children is not likely to be fault-prone than others with less number of children. The validation for both DIT and NOC are shown in Table 6

Table 6. DIT and NOC Validation

| Metric | Significant | Insignificant | N/A |
|--------|-------------|---------------|-----|
| DIT | [19],[22],[6],[25],[5],[2],[27], [9],[37] | [10],[21],[17],[23],[3],[4], [8],[24],[11],[29],[12],[30], [31],[32],[36] | [1],[28],[35],[34],[33] |
| NOC | [17],[22],[2] | [19],[25],[23],[3],[4],[24], [27],[11],[29],[12],[30],[31], [32],[36],[37] | [10],[21],[1],[6],[5],[2],[28],[9], [35],[34],[33] |

**N/A= not applicable

*Class Size measure:* In this study, analysis indicates that SLOC of a class has a strong relationship with FP and even more than OO metrics [1][17][22][25]. Consequently, about 17 studies confirmed its significance on FP and no study considered it insignificant while 12 studies did not measure it. (See Table 7) The implication of the results is that a class having a larger number of lines of code is more likely to have faults than classes with small code lines.

Table 7. SLOC Validation

| Metric | Significant | Insignificant | N/A |
|--------|-------------|---------------|-----|
| SLOC | [17],[1],[22],[25],[3],[4],[2],[24],[27], [28],[11],[29],[30],[31],[32],[33],[36] | - | [10],[19],[21],[6],[23],[8],[5],[12], [35],[34],[37] |

**N/A= not applicable

## 4.2. Empirical Validation Techniques

*RQ2: What techniques are being used to empirically validate the metrics in RQ1 and which is the best?* From the results of the analysis conducted, this study found that all the 29 studies selected explicitly stated the techniques used in conducting their individual empirical validation. Table 8 shows the techniques used, metrics studied, the variables employed (dependent and independent) and the tools employed for metric collection. However, different techniques were employed such as machine learning, logistic regression (LR) and so on. Moreover, LR is the most reported techniques used to construct predictive model that validate the relationship between metrics and FP. With these findings, we can deduced that LR is the best and widely used statistical techniques for predicting FP of a class CK+SLOC. Based on the analysis, about 76% of the studies used LR model (i.e. univariate and multivariate), and other 24% is for other techniques. (See Fig.1)

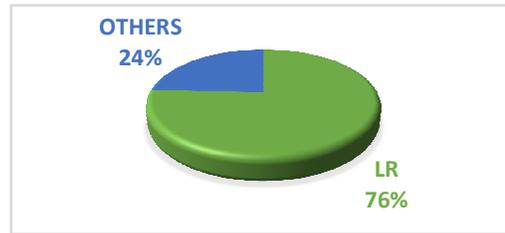

Figure 1. Statistical techniques used

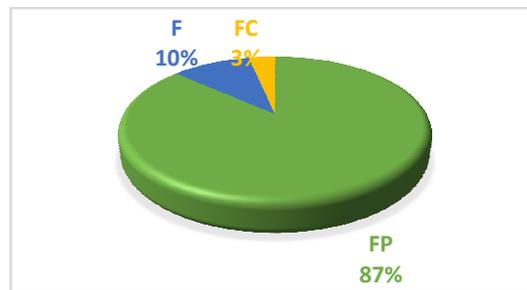

Figure 2. Used dependent variables used

Also, the variables used in the models are the dependent and independent variables which can be explained in terms of cause and effect. In an experiment, an independent variable is the cause or input, while the dependent variable is the output or effect [39]. To this end, dependent and independent variables are tested to validate if they are the actual effect and cause respectively by the prediction model. However, FP are used as the dependent variable in majority of the studies (87%) and 10% used faults data (F), while only 3% used fault count (FC) as dependent variable.(See Fig. 2) For independent variable, CK and SLOC metrics and others were specifically used.

Furthermore, metric collection method is considered to be critical to the accuracy of the metrics validated. From the analysis performed, it indicates that only 4% of the studies collected metrics manually, 41% stated the tools used in the collection, while 55% mentioned nothing about how metrics were collected. (See Fig. 3).

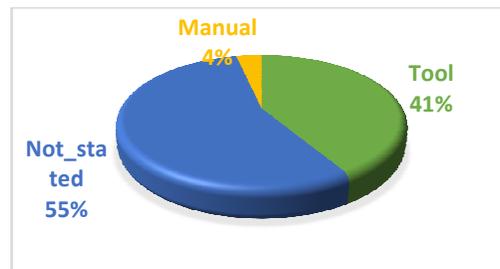

Figure 3. Metric collection methods

## 4.3. State of Metrics Validation

*RQ3: To what extent have the metrics in RQ1 been validated?* In this section, the state of the metric validation are considered from different points of views: *the study context*, *programming language used*, *product release time* and *the study type*. Table 8 present details of the metric validation state.

### 4.3.1. Study Subjects and Context

In this study based on the analysis conducted, it shows that the empirical validation studies of CK and SLOC metric's relationship with FP have been carried out in both academia and non-academia environments utilizing software products developed by either students or software professionals respectively. The academic environment used mainly systems developed by students while in the non-academia environment, either open source software (OSS) projects or industrial software systems developed by professionals were utilized. In most of the selected studies, product are either applications, components or middlewarethat ranges from OSS projects like Mozilla [23][25][28], eclipse [12][30][33], NASA project [1][3][4][11][31][36] to telecommunication systems [9][17][21]. Moreover, the systems have variable sizes ranging from small to large sized systems.

Table 8. Validation details

| Paper Id | Technique | Dependent Variable | Independent Variable | Metric Collection Tool | Prog. Language | Study Type | Release |
|---|---|---|---|---|---|---|---|
| 1 | LR | FP | CK & Others (24 ) | Java static analysis tool | JAVA | NR | Pre |
| 2 | LR | FP | CK  Metrics | M-System based on GEN++ | C++ | NR | Pre |
| 3 | LR | FP | CK & Others | commercial metrics collector | C++ | NR | Pre |
| 4 | OLS,LDA | FP | CK OTHERS 8 | Metric tool integrated with Rigi, | JAVA | NR | Pre |
| 5 | OLS,ANFIS | F | CK SLOC | - | C++ | NR | Pre |
| 6 | LR | FP | CK & OTHERS 49 | M-System based on GEN++ | C++ | NR | Pre |
| 7 | LR | FP | CK | M-System based on GEN++ | C++ | NR | Pre |
| 8 | LR/ML | FP | CK SLOC | Columbus | C++ | NR | Pre |
| 9 | LR | FP | CK OTHERS | Software System Markup Language | JAVA | NR | Pre |
| 10 | LR/ML | FP | CK SLOC | - | C++ | NR | Pre |
| 11 | LR/ML | FP | CK SLOC | - | C++ | NR | Pre |
| 12 | LR | FP | CK-NOC | - | C++ | NR | Pre |
| 13 | PRM, NBRM,ZINBRM | F | CK LOC | WebMetrics | C++ | NR | Pre |
| 14 | OLS | F | WMC,CBO,DIT,SIZE | - | C++, Java | NR | Pre |
| 15 | LR | FP | CK & OTHERS 49 | Manual | Java | R | Pre |
| 16 | LR | FP | CK & OTHERS 49 | A tool based on the FAST parser technology | C++ | R | Pre |
| 17 | LR | FP | WMC, LOC, Complexity | Software System Markup Language | JAVA | NR | Pre |
| 18 | ML (SVM) | FP | CK SLOC | - | C++ | NR | Pre |
| 19 | LR | FP | CK, LOC & OTHERS | - | C++ | NR | Pre |
| 20 | LR | FP | CK & OTHERS | Borland Together | JAVA | NR | Post |
| 21 | PRM,NBRM, ZINBRM | FP | CK & Others | - | C++ | NR | Pre |
| 22 | LR | FP | CK & Others | - | java | NR | Pre |

| 23 | LR | FP | CK & Others | - | C++ | NR | Pre |
| 24 | LR | FP | CK, LOC & OTHERS | - | java | NR | Pre |
| 25 | LR | FP | LCOM(CK) & others | - | java | NR | Pre |
| 26 | LR | FP | LCOM(CK) & others | - | java | NR | Pre |
| 27 | LR | FP | WMC, LOC, Complexity | - | JAVA | R | Pre |
| 28 | PRM, NBRM | FP,FC | CK, SLOC | - | - | NR | Pre |
| 29 | LR | FP | CK | - | JAVA | NR | Pre |

*LR: Logistic Regression; *ML: Machine Learning; *OLS, Ordinary Least Square; *LDA: Linear Discriminant Analysis;
*PRM: Poisson Regression Model;  *NBRM: Negative Binomial Regression Model; *ZINBRM: Zeros-Inflated Negative
Binomial Regression Model; *ANFIS: Adaptive Neuro-Fuzzy Inference System *SVM: Support Vector Machine.
R: Replicated, NR: Non-replicated; Pre: Pre-release; Post: Post-release.

As presented in Table 8, further analysis has shown that empirical validations is higher, about 79% in the non-academic environment than only 21% of the validation occurred in the academia environment like the University of Maryland (UMLD) [3][19][22] and University School of Information Technology (USIT) [24] and others (See Fig. 4). Furthermore, 21% of the systems used were written by students and 79% by mainly software professional few studies were replicated studies that utilizes same data sets from public repository such as eclipse, NASA, and others. (See Fig. 4). Also, the analysis shows that, 20% of the projects studied are student's projects, 33% are OSS and 47% are non-OSS systems.(See Fig. 5)

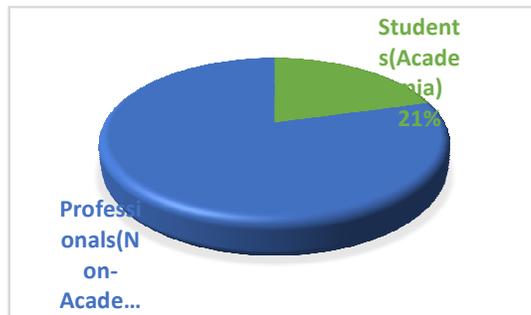

Figure 4. Validation environments

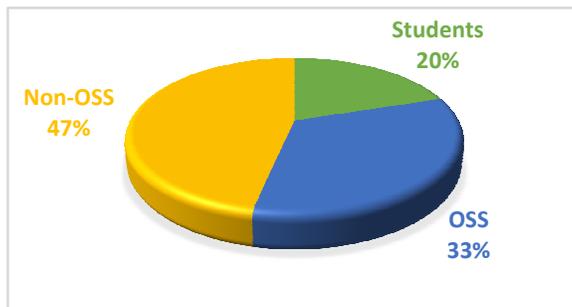

Figure 5. Projects validated

### 4.3.2. Programming Language Used

From the analysis carried out in this study, it indicates that only applications written with programming languages: Java and C++ were majorly used in the validation of the relationship between OO design metrics and FP. To this end, it shows that the two OO languages have dominated the world of software applications. However, analysis indicate that about 54% of the applications were written in C++ in both the industry and the academia while applications written in Java is about 43% and 3% of the studies did not mentioned the language of their application. (See Fig. 6)

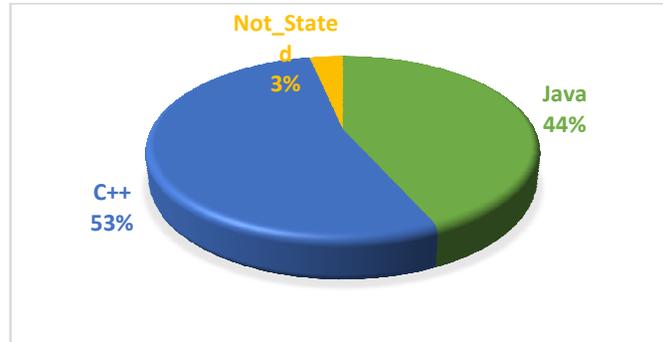

Figure 6.Programming languages used

### 4.3.3. Study Type and Product Release

In the context of this study, study type refers to whether the study is a replicated one or not. Replicated studies were considered in this study because only few studies exist on empirical validation of OO design metrics with respect to CK and SLOC while other studies were replicated. Basili et al [6] has stressed the need for replicated work as it assist to re-validate the metrics, provide understanding and usefulness of the metrics with regard to different types of faults. However, analysis shows that only 14% are replicated studies while 86% are non-replicated. This is captured in Fig. 7. Furthermore, Briand et al [22][27] replicated Basili et al[6], Aggarwal et al [24] replicated Briand et al [22][27] and Zhou et al [33] replicated Olague et al [23]. Other studies were also found reusing datasets of previous studies.

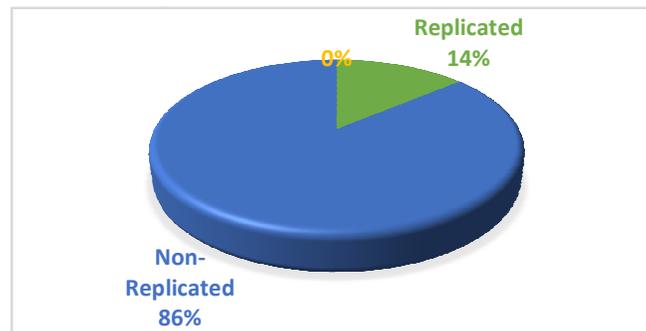

Figure 7. Study type

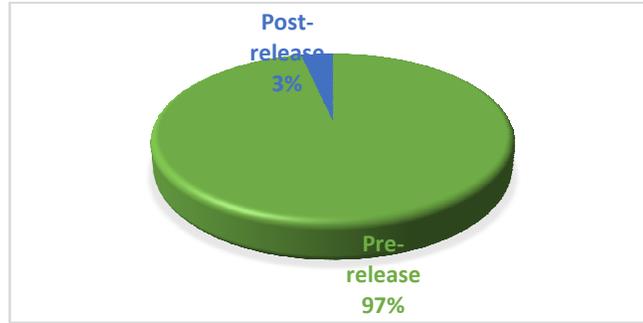

Figure 8. Product release type

Based on the release type, we mean the state of the system studied when its structural properties were measured and validated: pre-release and post-release. For instance, pre-release means measuring of faults during development and testing, while those faults measured after the system has been released to the users is the post-release. However, analysis indicates that about 97% of the systems used for the empirical validation where pre-release product. This findings confirmed the effectiveness of OO design metrics in evaluating the structural properties of OO classes. In addition, only 3% of the studies used post-release application (maintenance) by categorizing faults at different levels of severity (High, Medium and Low-impact errors) [12]. (See Fig. 8)

### 4.4.Metrics Empirical Validation Usefulness

*RQ4: Of what relevance are the empirical validations of software metrics?*In all the studies considered, it has been shown that empirical evidences is a vital step towards ensuring the practical relevance of software metrics in software organizations. It indicates that, without empirical evidence that product metrics are related to important external attributes like FP, metrics will remain little or of no value. In particular, studies by [2][10][16][21] provided an expression that depicts the theoretical basis for developing  prediction models for relating OO metrics and FP. The studies hypothesized that the relationship is due to the effects it has on cognitive complexity. (See Fig. 9) The indication is that the structural properties of classes have impact on cognitive complexity which in turn, relates to FP. More so, high cognitive complexity can lead OO classes exhibiting unwanted external qualities like FP, reduced understandability and maintainability [10]. Thus, metrics that having the ability to measure these structural properties are considered as good predictors of FP.

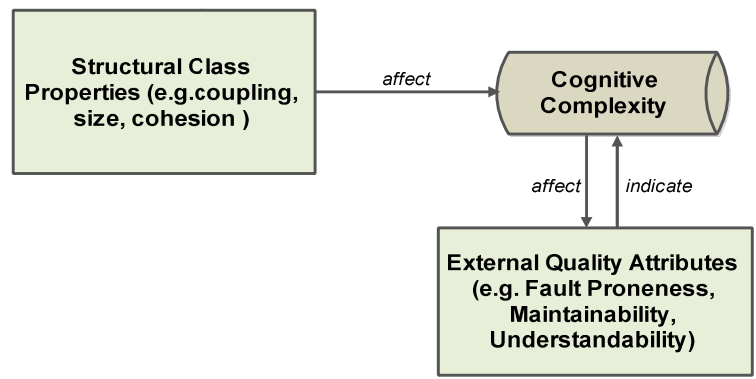

Figure 9. Theoretical basis of OO product metrics [10]

The studies went further to explain that the expression of such a relationship can be used for early prediction and identification of risky software classes or the construction of preventative (e.g. design, programming) strategies [10]. To this end, using OO design metrics such as CK and SLOC metrics can assist organizations to assess software development of any size swiftly at a reduced cost, take solution actions early and thus, avoid costly rework [10][11][19][24].

## 4.5. Generic Metric for Fault Proneness Prediction

*RQ5: Are there generic software metrics for predicting faulty classes?* In this SLR, analysis have shown that CK or CK and SLOC metrics have impact on class FP. Nonetheless, some studies did not consider some of the metrics. Moreover, the results are contradicting even when same dataset was used. For instance, in the study performed by [2] that utilized two systems written in C++ and Java, the results obtained indicates that WMC was significant with C++ but was not significant with Java. Also, DIT was significant in few studies but insignificant in most studies. This also applicable to other metrics. Fig. 10 presents the significance and insignificance distribution of CK and SLOC metrics on FP of OO classes.

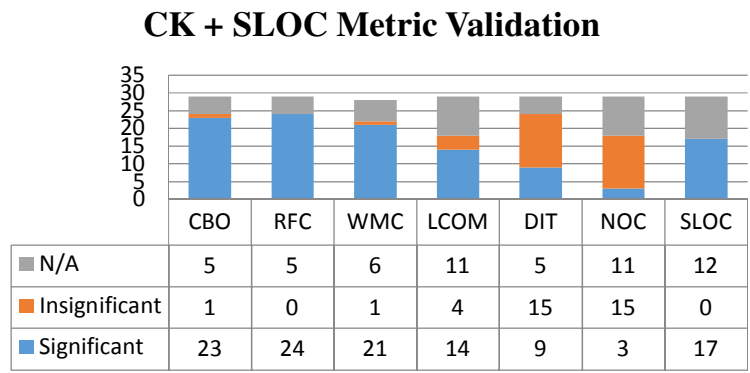

**CK + SLOC Metric Validation**

| | CBO | RFC | WMC | LCOM | DIT | NOC | SLOC |
|---|---|---|---|---|---|---|---|
| N/A | 5 | 5 | 6 | 11 | 5 | 11 | 12 |
| Insignificant | 1 | 0 | 1 | 4 | 15 | 15 | 0 |
| Significant | 23 | 24 | 21 | 14 | 9 | 3 | 17 |

Figure 10. Validation of CK + SLOC relationship with FP

From the results obtained in this analysis, it is clear that there is no generic metric for FP, rather best predictors of FP varies according to the type of applications used, the language used in coding and the target application domain. In addition, SLOC, CBO, RFC, and WMC are the metrics mostly reported as having significant relationship with FP in all the studies followed by LCOM. This confirms the findings in [40][41]. In this case, the results were based on the value of each metrics. Consequently, the higher the value, the higher the FP of the class. Moreover, DIT and NOC were the metrics found to be mostly insignificant in all the studies considered.

## 5. DISCUSSIONS

As OO programming has becomes the mainstream in software development today, several OO metrics have been proposed and developed to assess the quality of OO software systems. By assessing the quality of software during software development, quick design decisions at a reduced cost can be ensured. With the 29 studies considered in this SLR, it shows that only few empirical validation studies exist in the perspective of CK and SLOC metrics and FP prediction. However, the studies considered used different OO measures such as coupling, cohesion, inheritance and size measures to construct quality models that predicts the FP based on the statistical techniques of LR, machine learning and so on. In addition, the predictive accuracy of such models were reported based on either cross validation or goodness of fit [42]. Based on the analysis conducted, LR is the most widely used model with high predictive accuracy as well as the best in predicting faulty classes. These models utilizes FP as the dependent variable obtained during the testing phase, while the OO metrics are the independent variables obtained during design and coding phases. However, the statistical technique like LR can only predict the FP of a class without giving information regarding the possible number of faults in that class.

Also, the study found that size, complexity, coupling measures were the metrics found to be strongly related to FP followed by cohesion in the studies that considered CK+SLOC metrics. Inheritance measures were found to be insignificant in several studies. This led some authors to argued that DIT has an impact on the understandability of OO application and does not support reusability, while others argued that the number of methods in the classes is the factor that affects understandability [10][29]. With replicated studies, analysis shows that only few studies exist and most of the studies were based on shared or reused dataset of previous studies obtained from NASA, OSS (Mozilla, eclipse projects) and so on. Furthermore, results obtained from these studies were not consistent in terms of significance level. Some metrics appears to be significantly (positively or negatively) related to FP and some were not. Consequently, the best predictors of FP depends on the type of language, applications and the targeted domain.This study also found that the systems used in the empirical validation circled within the sphere of students, OSS, and non-OSS projects which is the dominant of all. In addition, majority of the systems were developed by professionals (79%). Also, validation were performed on only pre-release products (97%) and only one (4%) study actually performed it on post-release product. However, the study by [29] recommend that as a system evolves, it becomes more cumbersome to use OO metrics to accurately identify the FP of classes in post-release products. To this end, alternative methods needs to be applied if high accuracy is to be achieved. More so, only applications written in C++ and Java were used to validate the relationship between OO metrics and FP.

The implication of this study is that empirical validation of OO metrics relationship with FP is crucial to preserving the practical relevance of OO metrics in organizations. It can assist in the quick allocation of resources to where they are needed most, avoid the costly rework and facilitate other development activities such as change impact analysis, testing and so on. Therefore, during development strong efforts have to be technically channelled to keeping all those metrics at a reasonable level since FP of a class is based on each metric value.

## 5.1. Strengths and Weaknesses

This study covered at least large number of articles that assist in extracting relevant information used. To this end, we are quite sure that the study actually covers the empirical validation of CK and SLOC metrics published between January 1995 and December 2012. The SLR carefully followed the guidelines by proffered by Kitchenham et al [26] where credible and trusted sources were used. However, possible threats to this study could emanate from the search terms used, the risks posed by not covering all the relevant studies or it could be that most relevant studies were hidden in the excluded sources. Furthermore, threats could be the risk of misrepresenting the findings of some of the papers found like not considering fault severity levels, positive or negative significance of the metrics. Nonetheless, we have strong confidence that if such threats exist, they have no significant effect on the results of this SLR. In this case, we worked collaboratively, analysed all selected studies and all decisions as well as results were checked, rechecked and inconsistencies resolved.

## 6. CONCLUSIONS

Today, as the OO paradigm has gained widespread popularity coupled with software dependability, it is important that high software quality should not be compromised. OO design metrics should always be used to assess software quality during software development. By this evaluation, design quality will be improved which in turn would lower the probability of the software being flawed. Doing this at the early phases of development can attracts a considerably small cost and reduced efforts than late during development process. Several OO metrics have been proposed in this direction like CK metric suite and the size measure. Moreover, many empirical validation of the relationship between OO metric and FP have been reported. However, to ascertain which of them are useful predictors of FP, this study explored the existing empirical validation of CK+SLOC metrics to bring together the state-of-the-art in fault prediction using FP and CK + SLOC metrics. The results obtained were presented and discussed.

The main findings of this SLR are as follows:

- SLOC, CBO, RFC, WMC are metrics that strongly association with FP. There are also considered the best predictors of FP in majority of the studies. Moreover, LCOM is somehow an indicator of FP while DIT and NOC are found to be mostly insignificant. With the results, we deduced that best predictors of FP depends on the class of applications and the domain involved.
- This study found 29 empirical studies that have validated CK and SLOC metrics with FP of OO class. In these studies, 6 were from student's project and 23 were from non-students projects (mainly OSS and industrial applications).
- Software applications written in C++ and Java were majorly used to empirically validate the association between OO metrics and FP.
- The prediction models constructed were mainly based on LR. Only few machine learning and other techniques have been used. Thus, this study deduced that LR is the best statistical technique used for FP prediction.
- The empirical studies revolved around pre-release software products. Only one study has performed empirical validation on post-release product.
- Lastly, only few replicated studies exist. However, most studies were found reusing the dataset of previous studies.

Future work will involve conducting systematic review on the empirical validation of the relationship between FP and other OO metrics other than CK metric suite as well as maintainability.

With the above findings, here are some recommendations:

a) To predict the FP with some level of accuracy using CK and SLOC metrics, SLOC, CBO, RFC, WMC and LCOM are to be considered. Moreover, LR should be used as the predictive model. Metrics such as DIT, and NOC should only be considered based on the current value measured in that particular software product. This is because, though they appears not to be regular FP indicators, however their significance or insignificance could be as a result of either the developers' experience or the inheritance strategy applied.
b) For high quality software to be ensured that is stable and maintainable, low-coupling, highly cohesion, controlled size and inheritance should be adhered to.
c) For the evaluation of software quality during development or maintenance, measures should strongly not be based on the nature of the environment involved, instead on steady indicators of design problems and impacts on external quality attributes.
d) More empirical studies should be carried out on applications written in other OO languages other than C++ or Java. Also, additional empirical studies should be performed in the academia and more replicated studies should be carried out in order to re-validate the metrics and keep them relevant.
e) More efforts should be channeled towards post-release software products in order to confirm if models utilizing OO metrics can effectively predict class FP accurately or not.

f) During impact analysis of OO software systems, as a quality support activity, OO metrics can be used to assess the software quality first before actual changes are made.

To this end, developers and maintainers should use these metrics consistently to evaluate and then identify which OO classes requires attention in order to channel resources to those classes that are likely failure in the field.

## AUTHORS

**Dr. Isong, Bassey**

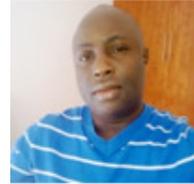

Received B.Sc. degree in Computer Science from the University of Calabar, Nigeria in 2004 and M.Sc. degrees in Computer Science and Software Engineering from Blekinge Institute of Technology, Sweden in 2008 and 2010 respectively. Moreover, he received a PhD in Computer Sciencein the North-West University, Mafikeng Campus, South Africa in 2014. Between 2010 and 2014 he was a Lecturer in the Dept. of Computer Science and Information Systems,University of Venda, South Africa. Currently, he is a Lecturer in the Department of Computer Sciences, Mafikeng Campus, North-West University. His research interests include Software Engineering, Requirements Engineering, Software Measurement, Maintenance, Information Security, Software Testing, Mobile Computing and Technology in Education.

**Prof. Obeten, Ekabua**

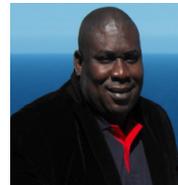

He is a Professor and Departmental Chair of the Department of Computer Science in the Delta State University, Abraka, Nigeria. He holds BSc (Hons), MSc and PhD degrees in Computer Science in 1995, 2003, and 2009 respectively. He started his lecturing career in 1998 at the University of Calabar, Nigeria. He is the former chair of the Department of Computer Science and Information Systems, University of Venda and Department of Computer Science, North-West University, Mafikeng Campus, South Africa. He has published several works in several journals and conferences. He has also pioneered several new research directions and made a number of landmarks contributions in his field and profession. He has received several awards to his credit. His research interest is in software measurement and maintenance, Cloud and GRID computing, Cognitive Radio Networks, Security Issues and Next Generation Networks.